\documentclass[proof]{WileyASNA-v1}
\articletype{Article Type}%
\received{14 January 2020}
\revised{xxx}
\accepted{xxx}

\begin{document}

\title{Properties of the QGP created in heavy-ion collisions}

\author[1]{P. Moreau} 
\author[2]{O. Soloveva}
\author[2]{I. Grishmanovskii}
\author[3]{V. Voronyuk}
\author[2]{L. Oliva}
\author[4]{T. Song}
\author[3]{V. Kireyeu}
\author[4]{G. Coci}
\author[4,2]{E. Bratkovskaya* }

\authormark{P. Moreau at al.}

\address[1] {Department of Physics, Duke University, Durham, NC 27708, USA}

\address[2]{\orgdiv{Institute for Theoretical Physics},
\orgname{ University of Frankfurt}, \orgaddress{\state{Frankfurt},  \country{Germany}}}

\address[3]{Joint Institute for Nuclear Research, Joliot-Curie 6, 141980 Dubna, Moscow Region, Russia}

\address[4]{\orgdiv{GSI}, \orgname{Helmholtzzentrum f\"{u}r Schwerionenforschung GmbH}, \orgaddress{\state{Darmstadt}, \country{Germany}}}

\corres{* GSI, Darmstadt, Germany \email{E.Bratkovskaya@gsi.de}}

\abstract{
We review the properties of the strongly interacting quark-gluon plasma (QGP) 
at finite temperature $T$ and baryon chemical potential $\mu_B$ as created in heavy-ion collisions at ultrarelativistic energies.
The description of the strongly interacting (non-perturbative) QGP in equilibrium 
is based on the effective propagators and couplings from the Dynamical QuasiParticle 
Model (DQPM) that is matched to reproduce the equation-of-state of the partonic system 
above the deconfinement temperature $T_C$ from lattice QCD. 
Based on a microscopic transport description of heavy-ion collisions
we discuss which observables are sensitive to the QGP creation and its properties.
}


\jnlcitation{\cname{%
\author{P. Moreau et al.}} (\cyear{2021}), 
\ctitle{Properties of the QGP created in heavy-ion collisions}, 
 \cjournal{Astronomische Nachrichten}, \cvol{2021;}.}

\maketitle

\section{Introduction}
An understanding of the structure of our universe
is an intriguing topic of research in our Millennium, which combines the
efforts of physicists working in different fields of astrophysics,
cosmology and heavy-ion physics \citep{Proceedings:2019wjy}. The common theoretical efforts and 
modern achievements in experimental physics allowed to make a substantial progress in understanding the properties of the nuclear matter and extended our 
knowledge of the QCD phase diagram which contains the information about the properties of our universe from the early beginning -- directly after the Big Bang -- when the matter was in a quark-gluon plasma (QGP) phase at very high temperature $T$ and practically zero baryon chemical potential $\mu_B$,
to the later stages of the universe, where in the expansion phase stars and galaxies have been formed. In the later phase, the matter is at low temperature, however, at very large baryon densities or baryon chemical potential $\mu_B$. The range of the phase diagram
at large $\mu_B$ and low temperature $T$  can also be explored in the astrophysical context  \citep{Klahn:2006ir}, such as in the dynamics of supernovae or in the dynamics of neutron-star mergers and gravitational waves.
Collisions of heavy-ions at ultra-relativistic energies -- 
e.g. at the Relativistic Heavy Ion Collider (RHIC) or the Large Hadron Collider (LHC) --  
provide the possibility to reproduce on Earth the conditions closer to the Big Bang,
when the matter was in a QGP phase of unbound quarks and gluons at very high temperature
and almost vanishing $\mu_B$. Indeed, in experiments it is possible to achieve a QGP only 
in extremely small volumes, moreover, the fast expansion leads to a fast hadronization of the QGP such that only final hadrons and leptons are measured.
Thus, it is quite challenging to investigate the properties of the QGP experimentally.

The region of phase diagram at finite $T$ and $\mu_B$ is of special interest nowadays.
According to lattice calculations of quantum chromodynamics
(lQCD)~\citep{Bernard:2004je,Aoki:2006we,Aoki:2009sc,Bazavov:2011nk,Gunther:2016vcp}, 
the phase
transition from hadronic to partonic degrees of freedom at
small baryon chemical potential $\mu_B \leq 350$ MeV is a crossover. 
According to  phenomenological models (e.g. recent PNJL calculations
\citep{Fuseau:2019zld}), the crossover is expected to turn into 
a first order transition at some critical point $(T_{cr}, \mu_{cr})$ in the phase diagram with increasing baryon chemical potential $\mu_B$. 
However, the theoretical predictions concerning the location of the critical 
point are rather uncertain. From the experimental side 
the beam energy scan (BES) program  at  RHIC  aims to find the
critical point and the phase boundary by scanning the 
collision energy~\citep{Mohanty:2011nm,Kumar:2011us}. Moreover,
new facilities such as FAIR (Facility for Antiproton and Ion
Research) and NICA (Nuclotron-based Ion Collider  fAcility) are
presently under construction; they will explore the intermediate energy
regime of rather large baryon densities and moderate temperatures 
where one might also study the competition between chiral symmetry restoration and 
deconfinement as advocated in Refs. \citep{Cassing:2015owa,Palmese:2016rtq}.

From the theoretical side it is quite challenging to investigate the "intermediate" 
region of the QCD phase diagram of non-vanishing quark (or baryon) densities 
from first principles due to the so called
"sign problem" which prevents the lQCD calculations to be extended to large $\mu_B$
(cf. \citep{Guenther:2020vqg}). Our present knowledge on QCD in Minkowski space 
for non-vanishing $\mu_B$ are based mainly on effective approaches.
Using effective models, one can study the properties of QCD in equilibrium, i.e., thermodynamic quantities, the Equation-of-State (EoS) of the QGP as well as transport coefficients. The dynamical quasiparticle model (DQPM)
has been introduced for that aim \citep{Peshier:2005pp,Cassing:2007nb,Cassing:2007yg,Linnyk:2015rco,Berrehrah:2016vzw}. 
It is based on partonic propagators with sizeable
imaginary parts of the self-energies incorporated. Whereas the real part of
the self-energies can be attributed to a dynamically generated mass (squared), the
imaginary parts contain the information about the interaction rates of 
the degrees-of-freedom. Moreover, the imaginary parts of the propagators define the spectral functions of the 'particles' which might show narrow (or broad) quasiparticle peaks.
An important advantage of a propagator based approach is that one can formulate a
consistent thermodynamics~\citep{Vanderheyden:1998ph} and
a causal theory for non-equilibrium states on the basis of Kadanoff--Baym equations~\citep{KadanoffBaym}.

Since the QGP is formed for a short time in a finite volume in heavy-ion collisions (HICs), 
it is very important to understand the time evolution of the expanding system. Thus, 
it is mandatory to have a proper non-equilibrium description of the entire dynamics through different phases - starting with impinging nuclei in their groundstates, 
going through  the QGP phase (showing some approach to equilibration) and 
end up with the final asymptotic hadronic states. 
To this aim, the Parton--Hadron--String Dynamics (PHSD) transport approach
~\citep{Linnyk:2015rco,Cassing:2008sv,Cassing:2008nn,Cassing:2009vt,Bratkovskaya:2011wp}
has been formulated more than a decade ago (on the basis of the
Hadron-String-Dynamics (HSD) approach~\citep{Cassing:1999es}),
and it was found to well describe observables from p+A and A+A collisions 
from SIS (Schwerionensynchrotron) to LHC energies for the bulk dynamics, 
electromagnetic probes such as photons and dileptons
as well as open cham hadrons \citep{Bratkovskaya:2004cq,Linnyk:2015rco,Linnyk:2013hta,Song:2018xca}.

In order to explore the partonic systems at finite $\mu_B$, the PHSD approach has been extended to incorporate partonic quasiparticles and their differential cross sections that depend not only on temperature $T$ as obtained in Ref. \citep{Ozvenchuk:2012fn}
and employed in the previous PHSD studies, but
also on chemical potential $\mu_B$ and center-of-mass energy of
colliding partons $\sqrt{s}$ and their angular distributions 
explicitly \citep{Moreau:2019aux,Moreau:2019vhw}.
Within this extended approach we have studied the `bulk' observables
such as rapidity distributions and transverse momentum spectra 
in heavy-ion collisions from AGS (Alternating Gradient Synchrotron) to RHIC energies for symmetric and asymmetric 
(light + heavy nuclear) systems. However, we have found only a small influence 
of $\mu_B$ dependences of parton properties (masses and widths) and partonic interaction
cross sections in the bulk observables~\citep{Moreau:2019vhw}.

Recently we studied  the collective flow ($v_1$, $v_2$) coefficients for different identified hadrons and their sensitivity to the $\mu_B$ dependences of partonic cross sections  \citep{Soloveva:2020xof,Soloveva:2020ozg}.
We have found that the flow coefficients show a small, but visible sensitivity
to the $\mu_B$ dependence of partonic interactions.  

In this work we study the properties of the QGP created in heavy-ion
collisions and show the time evolution of the $(T,\mu_B)$ distribution,
as probed in HICs, and relate it to observables such as multiplicities
and collective flow ($v_1$, $v_2$) coefficients.

\section{A microscopic transport description of the nonperturbative QGP}

In this Section we recall the basic ideas of the PHSD transport approach and the dynamical quasiparticle model (DQPM).
The~Parton--Hadron--String Dynamics (PHSD) transport approach \citep{Linnyk:2015rco,Cassing:2008sv,Cassing:2008nn,Cassing:2009vt,Bratkovskaya:2011wp}
is a microscopic off-shell transport approach for the description of strongly interacting hadronic and partonic matter in and out-of equilibrium. 
It is based on the solution of
Kadanoff--Baym equations in first-order gradient expansion in phase space
\citep{Cassing:2008nn} which allows to describe in a causal way the time evolution of nonperturbative interacting systems. Consequently they are applicable for the dynamics of the QGP which approximately behaves as a strongly interacting liquid at finite $T$ due to the
growing of the QCD coupling constant in the vicinity of the critical
temperature $T_C$, where pQCD methods are not applicable. 

The partonic phase in the PHSD is modelled based on the  dynamical quasiparticle model
(DQPM) \citep{Peshier:2005pp,Cassing:2007nb,Cassing:2007yg}.
The DQPM has been introduced in Refs.~\citep{Peshier:2005pp,Cassing:2007nb,Cassing:2007yg} for the effective description
 of the QGP in terms of strongly interacting quasiparticles - quarks and gluons,
where the properties and interactions are adjusted to reproduce lQCD results for 
 the QGP in equilibrium  at finite temperature $T$ and
baryon chemical potential $\mu_B$ \citep{Bernard:2004je,Aoki:2006we,Aoki:2009sc,Bazavov:2011nk,Gunther:2016vcp}.
In the DQPM, the quasiparticles are characterized by single-particle Green's functions (in propagator representation) with complex self-energies.
The real part of the self-energies is related to the dynamically generated parton masses (squared),
whereas the imaginary part provides information about the lifetime and/or
reaction rates of the degrees-of-freedom. Thus, in the DQPM the properties of the partons (quarks and gluons) are  characterized by broad spectral functions $\rho_j$ ($j=q, {\bar q}, g$), i.e. the partons are off-shell. This differentiates the PHSD from conventional cascade or transport models dealing with on-shell particles, i.e.,  $\delta$-functions in the invariant mass squared.
The quasiparticle spectral functions are assumed to have a Lorentzian form~\citep{Linnyk:2015rco}, which are specified by 
 the parton masses and width parameters:
\begin{eqnarray}
\!\!\!\!\!\! \rho_j(\omega,{\bf p}) =
 \frac{\gamma_j}{E_j} \left(
   \frac{1}{(\omega-E_j)^2+\gamma_j^2} - \frac{1}{(\omega+E_j)^2+\gamma_j^2}
 \right)\
\label{eq:rho}
\end{eqnarray}
separately for quarks/antiquarks and gluons ($j=q,\bar{q},g$). With the convention $E^2({\bf p}^2) = {\bf p}^2+M_j^2-\gamma_j^2$, the parameters $M_j^2$ and $\gamma_j$ are directly related to the real
and imaginary parts of the retarded self-energy, e.g.,~$\Pi_j =
M_j^2-2i\gamma_j\omega$.
The functional forms, i.e. the ($T, \mu_B$)-dependences of the
dynamical parton masses and widths are chosen in the spirit of
hard-thermal loop (HTL) calculations by introducing three parameters.
They are determined by comparison to the calculated entropy density $s$, pressure
$P$ and energy density $\epsilon$ from the DQPM to those from lQCD at
$\mu_B=0$ from Ref.~\citep{Borsanyi:2012cr,Borsanyi:2013bia}.

The DQPM also allows to define a scalar mean-field $U_s(\rho_s)$ for quarks and antiquarks which can be
expressed by the derivative of the potential energy density with respect to the
scalar density $\rho_s(T,\mu_B)$,
\begin{equation} \label{uss}
U_s(\rho_s) = \frac{d V_p(\rho_s)}{d \rho_s} ,
\end{equation}
which is evaluated numerically within the DQPM. Here, the potential energy density
is evaluated by:
\begin{equation} \label{Vp}
V_p(T,\mu_B) = T^{00}_{g-}(T,\mu_B) + T^{00}_{q-}(T,\mu_B) + T^{00}_{{\bar q}-}(T,\mu_B),
\end{equation}
where the different contributions $T^{00}_{j-}$ correspond to the
space-like part of the energy-momentum tensor component $T^{00}_{j}$
of parton $j = g, q, \bar{q}$ (cf. Section~3 in Ref.
\citep{Cassing:2007nb}). The scalar mean-field $U_s(\rho_s)$ for quarks
and antiquarks is repulsive as a function of the parton scalar
density $\rho_s$ and shows that the scalar mean-field potential is in the
order of a few GeV for $\rho_s > 10$ fm$^{-3}$. The mean-field potential
(\ref{uss}) determines the force on a partonic quasiparticle $j$, i.e.,~$ \sim
M_j/E_j \nabla U_s(x) = M_j/E_j \ d U_s/d \rho_ s \ \nabla
\rho_s(x)$, where the scalar density $\rho_s(x)$ is determined
numerically on a space-time~grid in PHSD.

The quasiparticles - quarks and gluons - which are moving in the self-generated 
scalar mean-field potential (\ref{uss}) can interact via elastic and inelastic
scattering.
A two-body interaction strength can be extracted from
the DQPM as well from the quasiparticle widths in line with Ref.~\citep{Peshier:2005pp}. In the latest version of PHSD (v. 5.0) the following elastic and inelastic
interactions are included 
$qq \leftrightarrow qq$, $\bar{q} \bar{q} \leftrightarrow \bar{q}\bar{q}$, $gg \leftrightarrow gg$,
$gg \leftrightarrow g$, $q\bar{q} \leftrightarrow g$, $q g \leftrightarrow q g$,
$g \bar{q} \leftrightarrow g \bar{q}$. The backward reactions are defined using
'detailed-balance' with cross sections calculated from the leading order Feynman diagrams employing the effective propagators and couplings $g^2(T/T_c,\mu_B)$ 
from the DQPM~\citep{Moreau:2019vhw}. In~Ref.~\citep{Moreau:2019vhw}, 
the differential and total off-shell
cross sections have been evaluated as a function of the invariant energy of the
colliding off-shell partons $\sqrt{s}$ for each $T$, $\mu_B$.
We recall that in the early PHSD studies (using version 4.0 and below) the cross sections depend only on $T$ (cf. the detailed evaluation in Ref.~\citep{Ozvenchuk:2012fn}).

In the PHSD5.0 the quasiparticle properties (masses and widths) as well as their interactions, defined by the differential cross sections and mean-field potential, depend on the 'Lagrange parameters'  $T$ and $\mu_B$ in each computational cell in space-time.  
The evaluation of $T$ and $\mu_B$, as calculated in the PHSD from the local
energy density $\varepsilon$ and baryon density $n_B$, is realized 
by employing the lattice equation of state and  and by a diagonalization of the energy-momentum tensor from PHSD as described in Ref. \citep{Moreau:2019vhw}.

The hadronization, i.e. the transition from partonic to hadronic degrees-of-freedom 
(and vice~versa)  is described by covariant transition rates for 
the fusion of quark--antiquark pairs or three quarks (antiquarks), respectively.
The full microscopic description allows to obey flavor
current--conservation, color neutrality as well as energy--momentum
conservation~\citep{Cassing:2009vt}. Since the dynamical quarks and
antiquarks become very massive close to the phase transition, the formed resonant 'prehadronic' color-dipole states ($q\bar{q}$ or
$qqq$) are of high invariant mass, too, and sequentially decay to
the groundstate meson and baryon octets, thus increasing the total
entropy during hadronization.

On the hadronic side, PHSD includes explicitly the baryon octet and
decouplet, the~$0^-$- and $1^-$-meson nonets as well as selected
higher resonances as in the Hadron--String--Dynamics (HSD)
approach~\citep{Cassing:1999es}.  Note that PHSD and HSD
(without explicit partonic degrees-of-freedom) merge at low energy
density, in particular below the local critical energy density
$\varepsilon_c\approx$ 0.5~GeV/fm$^{3}$ as extracted from the 
lQCD results in Ref.~\citep{Borsanyi:2012cr,Borsanyi:2013bia}.

\section{The transport properties of the QGP close to equilibrium}

The transport properties of the QGP close to equilibrium are characterized by
various transport coefficients. The transport coefficients 
play an important role for the hydrodynamic models since they define 
the properties of the propagating fluid.
The shear viscosity $\eta$ and bulk viscosity $\zeta$ describe the fluid's
dissipative corrections at leading order. 
In the hydrodynamic equations, the viscosities appear as dimensionless ratios,
$\eta/s$ and $\zeta/s$, where $s$ is the fluid entropy density.
Such specific viscosities are more meaningful than the unscaled $\eta$ and $\zeta$
values because they describe the magnitude of stresses inside
the medium relative to its natural scale.
Both coefficients $\eta$ and $\zeta$ are generally expected 
to depend on the temperature $T$ and baryon chemical potential $\mu_B$.
While for the hydrodynamic models the transport coefficient are an 'input'
and have to be adopted from some models, a fully miscoscopic description
of the QGP dynamics by transport approaches allows to define them 
from the interactions of the underlying degrees of freedom.

One way to evaluate the viscosity coefficients of partonic matter is the Kubo formalism~\citep{Kubo:1957mj,Aarts:2002cc,FernandezFraile:2005ka,Lang:2015nca}, which was used to calculate the viscosities for a previous version of the DQPM within the PHSD transport approach in a box with periodic boundary conditions in Ref.~\citep{Ozvenchuk:2012fn} as well as in the more recent study with
the DQPM model in Refs.~\citep{Moreau:2019vhw,Soloveva:2019xph}.
Another way to calculate transport coefficients  is to use the relaxation--time approximation (RTA)
as incorporated in Refs. \citep{Hosoya:1983xm,Chakraborty:2010fr,Albright:2015fpa,Gavin:1985ph}. This strategy has been used in our recent studies where  we have
investigated the transport properties of the QGP in the $(T,\mu_B)$ plane based on the DQPM \citep{Moreau:2019vhw,Soloveva:2019xph,Soloveva:2019doq}.

We find that the ratios of transport coefficients $\eta/s$ and $\zeta/s$ evaluated within DQPM 
are in a good agreement with the available lQCD results for pure SU(3) gauge theory ~\citep{Astrakhantsev:2017nrs,Astrakhantsev:2018oue}. The $\eta/s$ ratio shows a
minimum at $T_C$ and slightly rises with $T$, while $\zeta/s$ grows at $T_C$
in line with the lQCD data. This shows that the QGP
in the PHSD behaves as a strongly interacting nonperturbative system 
rather then a dilute gas of weakly interacting partons.

\section{Properties of QGP in Heavy-Ion~Collisions}

\begin{figure*}[h!]
	\centering
	\includegraphics[width=10cm]{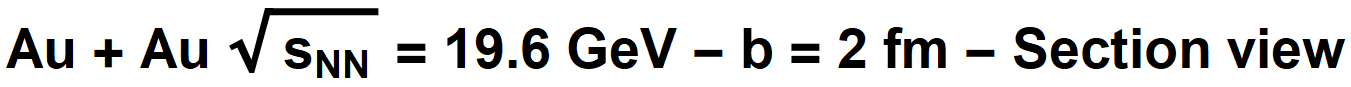} 	~\\ \vspace{0.2cm}
\hspace*{5mm}	\includegraphics[width=16cm]{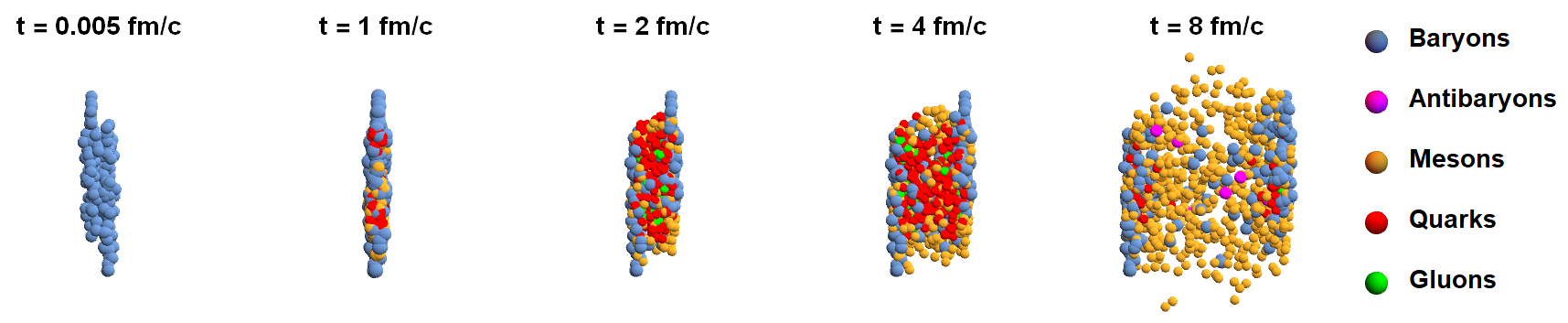} 	~\\ \vspace{0.2cm}
	\includegraphics[width=15cm]{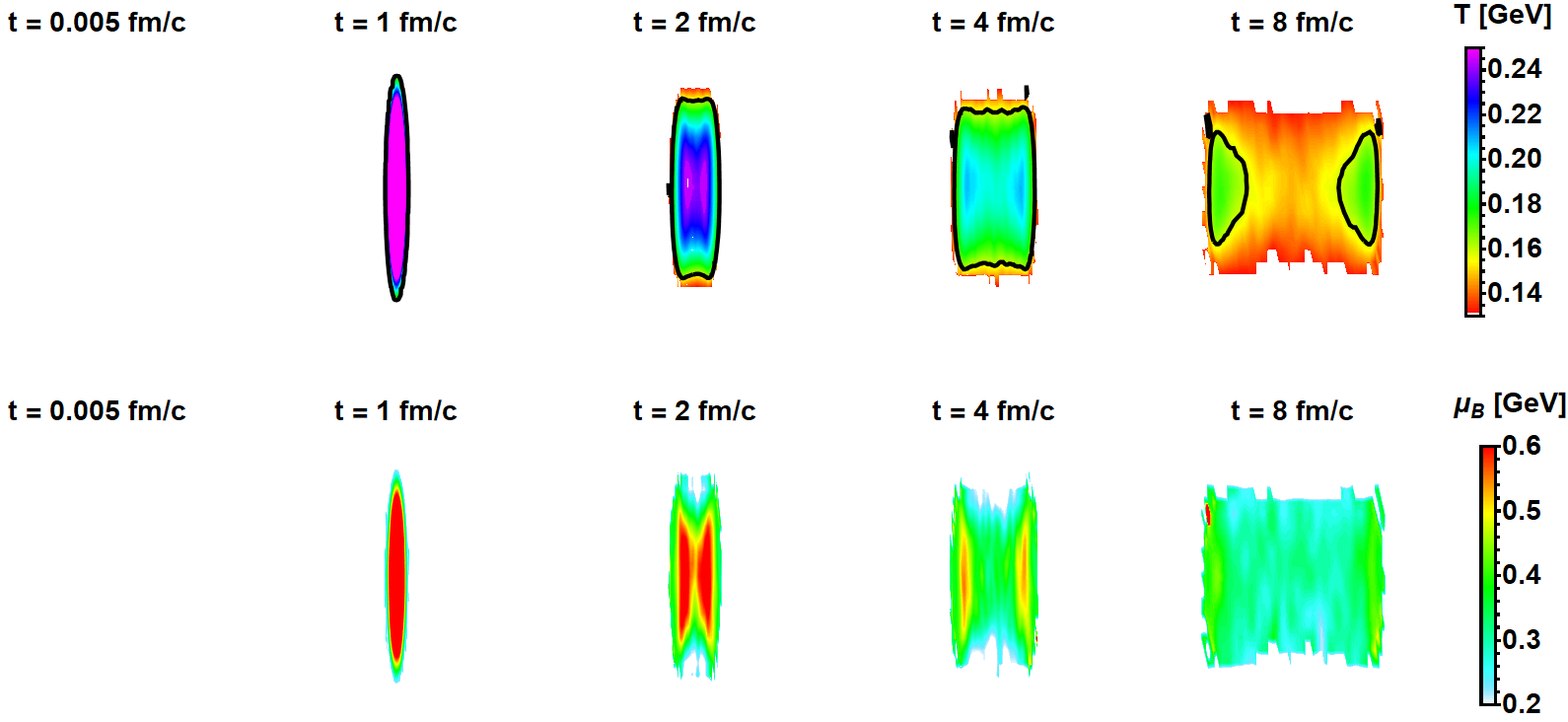}		
\caption{Illustration of the time evolution of central Au+Au collisions (upper row, section view) at a collisional energy of $\sqrt{s_{NN}} = 19.6$ GeV within the PHSD 
\citep{Moreau:2019aux}. 
The local temperature $T$ (middle row), baryon chemical potential $\mu_B$ (lower row),  as extracted from the PHSD for $y \approx 0$. 
The black lines (middle row) indicate the critical temperature $T_c = 0.158$ GeV.}
	\label{Fig_HIC}
\end{figure*}

In this Section we discuss the properties of the QGP created in 
heavy-ion collisions in terms of achievable energy densities and temperatures
as well as baryon chemical potentials. 
We start with an illustration in Fig. \ref{Fig_HIC} of a time evolution 
of central Au+Au collisions (upper row, section view) at a collisional energy 
of $\sqrt{s_{NN}} = 19.6$ GeV within the PHSD \citep{Moreau:2019aux}. 
The snapshots are taken at times $t=0.005, 1, 2, 4$ and 8 fm/c.
The baryons, antibaryons, mesons, quarks and gluons  are shown as colored dots.

We show the local temperature $T$ (middle row) and baryon chemical potential 
$\mu_B$ (lower row), as extracted from the PHSD in the region with $y \approx 0$. 
The black lines (middle row) indicate the critical temperature $T_c = 0.158$ GeV.

As follows from the upper part of Fig. \ref{Fig_HIC}, the QGP is created 
in the early phase of collisions and when the system expands,  hadronization occurs.
One can see that during the overlap phase the $T$ and $\mu_B$ are very large
(we give a warning that an extraction of the thermodynamic quantities 
for the strongly non-equilibrium initial stage is not a consistent procedure)
and they decrease with time. However even at 8 fm/c there are "hot spots" of QGP
at front surfaces of high rapidity.


\begin{figure}[h!]
	\centering
		\includegraphics[width=6cm]{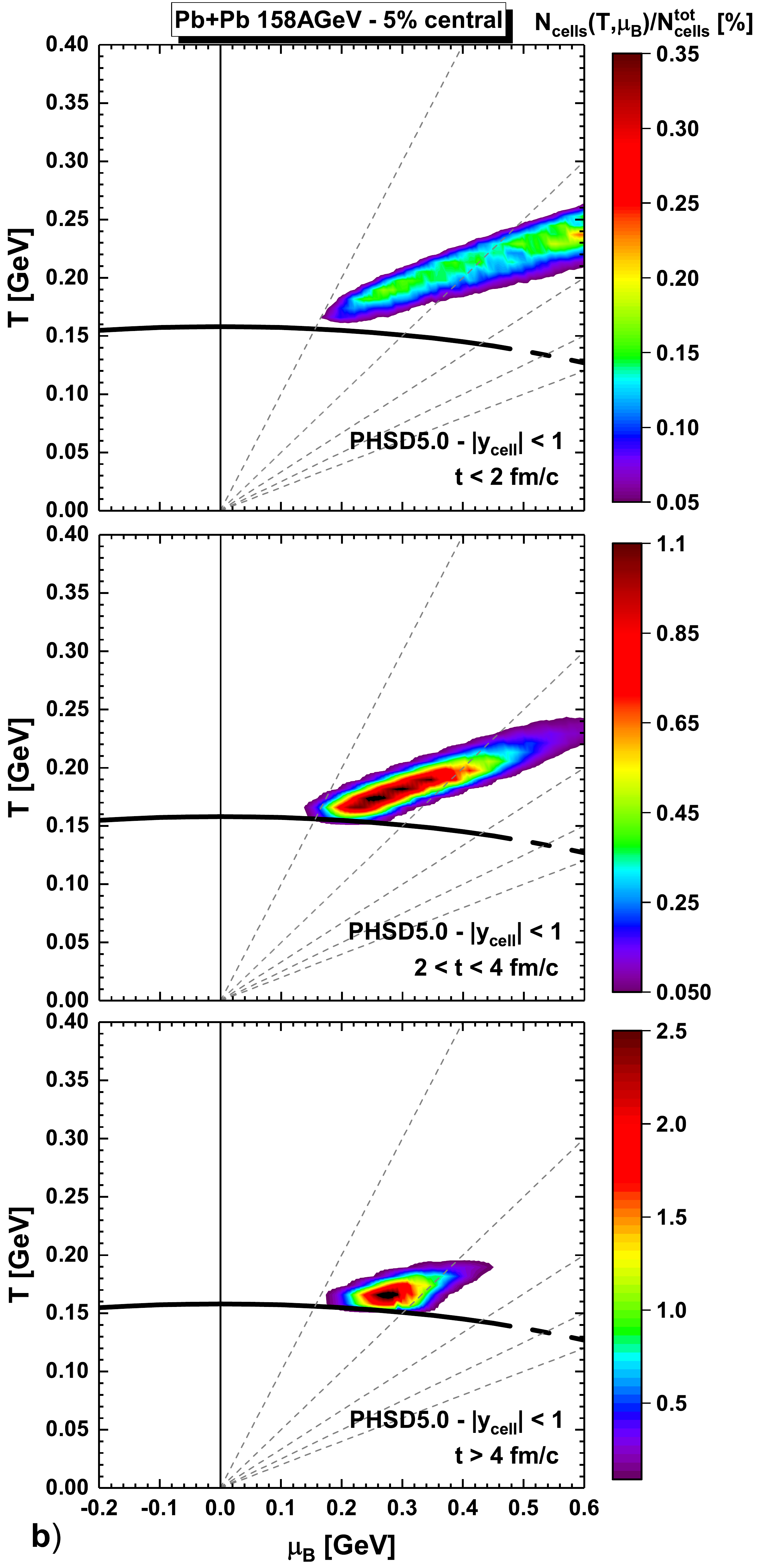}
	\caption{Distributions in $T$ and $\mu_B$ as extracted from the DQPM equation of state in a PHSD simulation of a central Pb+Pb collision at 158 A GeV  for cells with a temperature $T > T_c(\mu_B)$ at midrapidity ($|y_{\text{cell}}| <$ 1). The scale corresponds to the number of cells in the PHSD event in the considered bin in $T-\mu_B$ divided by the total number of cells in the corresponding time window (see legend). The solid black line is the DQPM phase boundary for orientation; the gray dashed lines indicate ratios of $\mu_B/T$ ranging from 1 to 5 while the vertical line corresponds to $\mu_B$ = 0.}
	\label{fig-Eval-T-muB-200GeV-158AGeV}
\end{figure}
We investigate now the time evolution of the $T$ and $\mu_B$ distribution in  for cells having a temperature $T > T_c(\mu_B)$ at midrapidity ($|y_{\text{cell}}| <$ 1) for 
5\% central Pb+Pb collisions at 158 A GeV.
Fig. \ref{fig-Eval-T-muB-200GeV-158AGeV} shows this distribution for times $t<$ 2 fm/c, 2 fm/c $<t<$ 4 fm/c and $t>$ 4 fm/c. For early times $t<$ 2 fm/c the distribution peaks at a temperature of about 0.25 GeV and a sizable chemical potential of about 0.6 GeV while for times in the interval 2 fm/c $ <t<$ 4 fm/c the maximum has dropped already to an average temperature $\sim$ 0.18 GeV and a chemical potential of about 0.3 GeV. For later times $t>$ 4 fm/c the distribution (above $T_c$) essentially stays around $\mu_B \approx 0.25$ GeV. We mention that the values of $\mu_B$ probed around the transition temperature $T_c$ in the PHSD are in accordance with the expectation from statistical models, which for central Pb+Pb collisions at 158 A\ GeV quote a value of $\mu_B = 0.2489$ GeV \citep{Cleymans:2005xv}.

\begin{figure}[h!]
	\centering
	\includegraphics[width=7.cm]{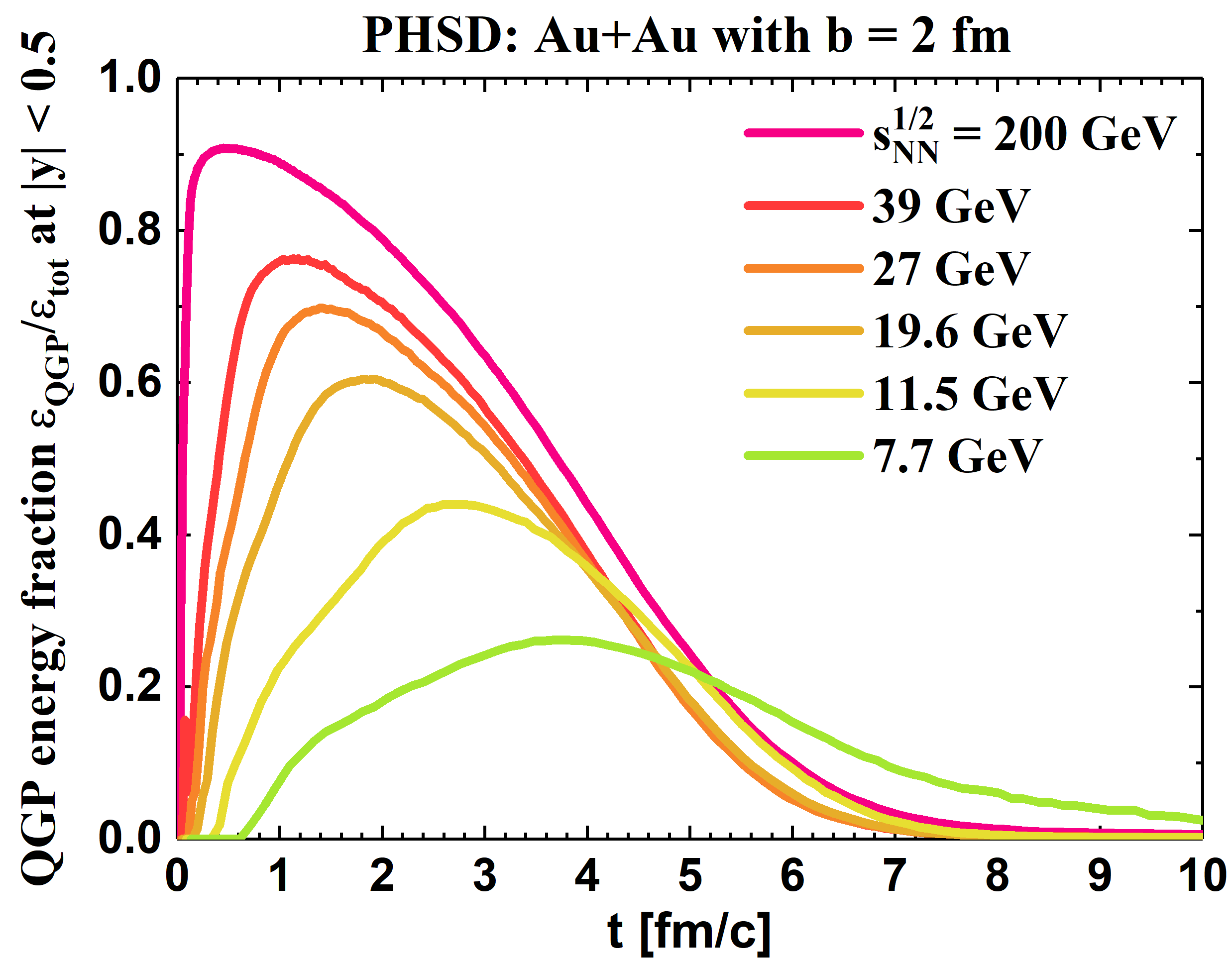} 
	\caption{The QGP energy fraction from PHSD  as a function of time $t$ in central (impact parameter b = 2 fm) Au+Au collisions for different collisional energies $\sqrt{s_{NN}}$, taking into account only the midrapidity region $|y| < 0.5$  \citep{Moreau:2019aux}.}
	\label{Fig_QGPfrac}
\end{figure}

By varying the collisional energy of the initial nuclei, one can increase or decrease
the volume of the QGP produced in the heavy-ion collisions.
In Fig. \ref{Fig_QGPfrac} we show the QGP energy fraction versus the total energy 
for Au+Au at different collisional energies $\sqrt{s_{NN}}$ accounting only the midrapidity region $|y| < 0.5$. One can see that for high energies the QGP fraction is  large compared to lower collisional energies where the volume of QGP is small. 
While at high energy heavy-ion collisions the QGP phase appears suddenly after 
the initial primary NN collision, at low energies it appearance is smoother 
since the passing time of the nuclei is longer. Correspondingly, at low energies the
QGP lifetime is large, however, the QGP volume is very small; thus it influence
on the dynamics is much reduced compared to high energy collisions where
practically 90\% of matter at midrapidity is in the QGP phase 
(at least for a short time).  

\begin{figure}[h!]
	\centering
	\includegraphics[width=7cm]{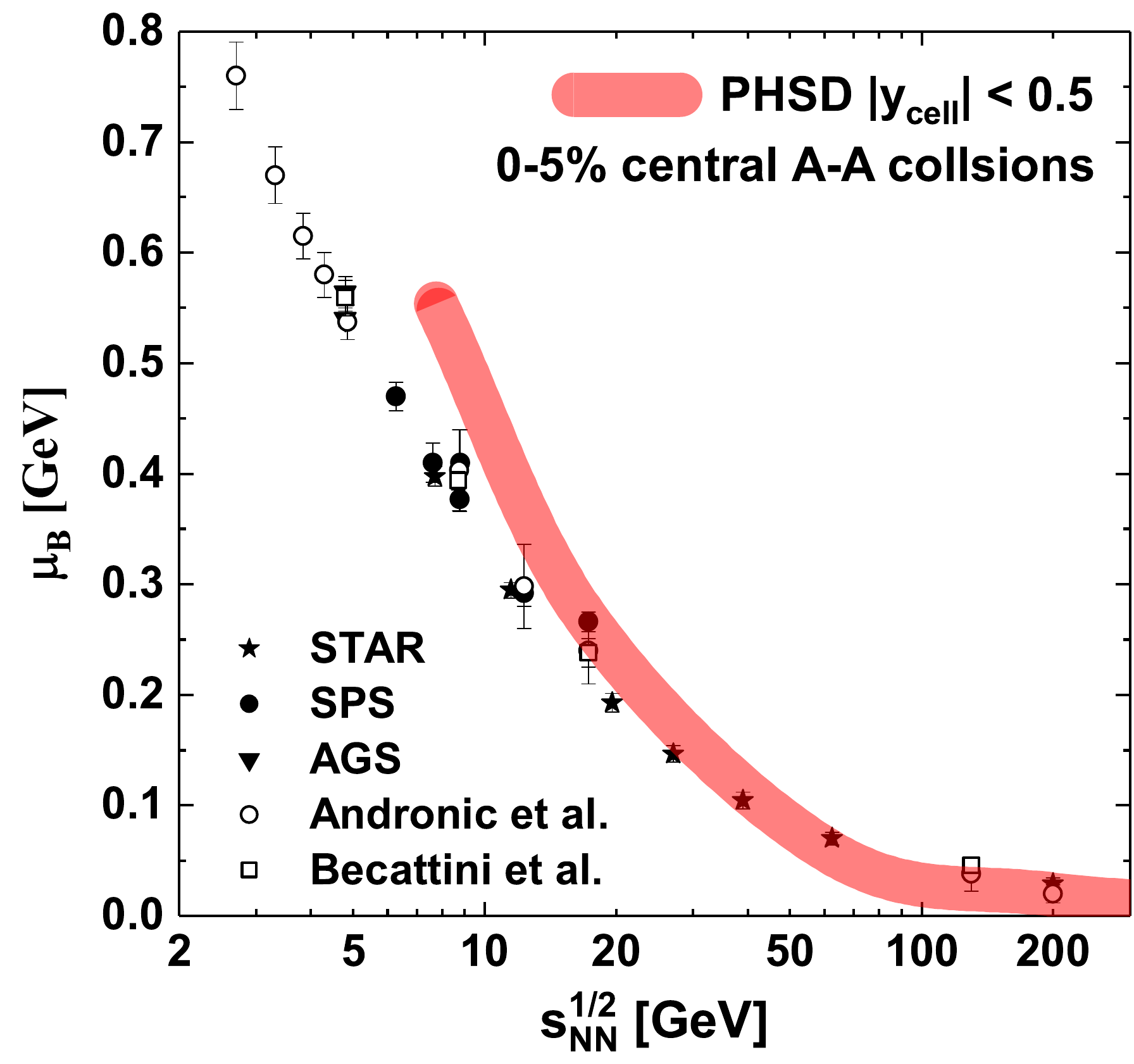}
\caption{ Average baryon chemical potential $\mu_B$ extracted from PHSD simulations \citep{Moreau:2019aux} around the chemical freeze-out temperature $T_{ch}$ 
in comparison with the values obtained from statistical models 
\citep{Andronic:2009jd,Cleymans:2005xv,Adamczyk:2017iwn}.}
	\label{fig-snB}
\end{figure}

In Fig. \ref{fig-snB} we show the average baryon chemical potential (from PHSD simulations) taken around the chemical freeze-out temperature (given by the statistical analysis from Refs. \citep{Andronic:2009jd,Cleymans:2005xv,Adamczyk:2017iwn}). We look only at cells in the midrapidity region to compare with the results from the statistical models. One should  stress that we compare two  different quantities here: one is $\mu_B$ obtained from a statistical analysis of the final particle spectra, and the other is the average $\mu_B$ probed in a PHSD simulation around the chemical freeze-out temperature (the latter itself being obtained by a statistical analysis). 
As one can see in Fig. \ref{fig-snB}, these two quantities are in a fairly good agreement especially at high energies where $\mu_B$ is rather small. At low collisional energies the extracted value of $\mu_B$ from the transport calculation (from cells at midrapidity) becomes slightly larger; this might partly be related to the fact 
that our $T, \mu_B$ extraction procedure, optimized for the QGP is
not sufficiently accurate for such large values of $\mu_B/T$ as probed in this regime.
The work on improving the $T,\mu_B$ extraction for hadronic dominated matter is in progress.

\section{Observables in Heavy-Ion~Collisions}

In our recent study~\citep{Moreau:2019vhw} we have investigated the sensitivity
of 'bulk' observables, such as rapidity and transverse momentum distributions
of different hadrons produced in symmetric and asymmetric heavy-ion collisions from AGS to top RHIC energies, on the details of the QGP interactions and the properties of the partonic degrees-of-freedom. For that, we have considered the following three cases:
(1) `PHSD4.0': the quasiparticle properties (masses and widths of quarks and gluons)
and partonic interaction cross sections depend only on $T$ as  calculated in Ref.~\citep{Ozvenchuk:2012fn}.
(2) `PHSD5.0 - $\mu_B$': the masses and widths of quarks and gluons depend on $T$
and $\mu_B$ explicitly; the differential and total partonic cross sections are obtained by calculations of the leading order Feynman diagrams from the DQPM and explicitly depend on invariant energy $\sqrt{s}$, temperature $T$ and baryon chemical potential $\mu_B$ ~\citep{Moreau:2019vhw}.
(3) `PHSD5.0 - $\mu_B=0$': the same at (2), but for $\mu_B=0$.

The comparison of the 'bulk' observables for A+A collisions within the three cases of PHSD in Ref.~\citep{Moreau:2019vhw} has illuminated that they show a very low sensitivity to the $\mu_B$ dependences of parton properties (masses and widths) and their interaction cross sections such that the results
from PHSD5.0 with and without $\mu_B$ were very close to each other.
Only in the case of kaons,  antiprotons $\bar{p}$ and antihyperons $\bar{\Lambda} + \bar{\Sigma}^0$, a small difference between PHSD4.0 and PHSD5.0 could be seen at
top SPS and top RHIC energies. 
This can be understood as follows: as has been illustrated in Section 4, 
at high energies such as top RHIC where the QGP volume is very large in central collisions, the~$\mu_B$ is very small, while, when decreasing the collision
energy (and consequently increase $\mu_B$) the fraction of the QGP is decreasing 
such that the final observables are dominated by the hadronic phase. Thus, the probability for the hadrons created at the QGP hadronization to rescatter, 
decay, or be absorbed in hadronic matter increases strongly; 
as a result the sensitivity  to the properties of the QGP is washed out to a large extend.

\begin{figure}[h!]
	\centering
	\includegraphics[width=9.5cm]{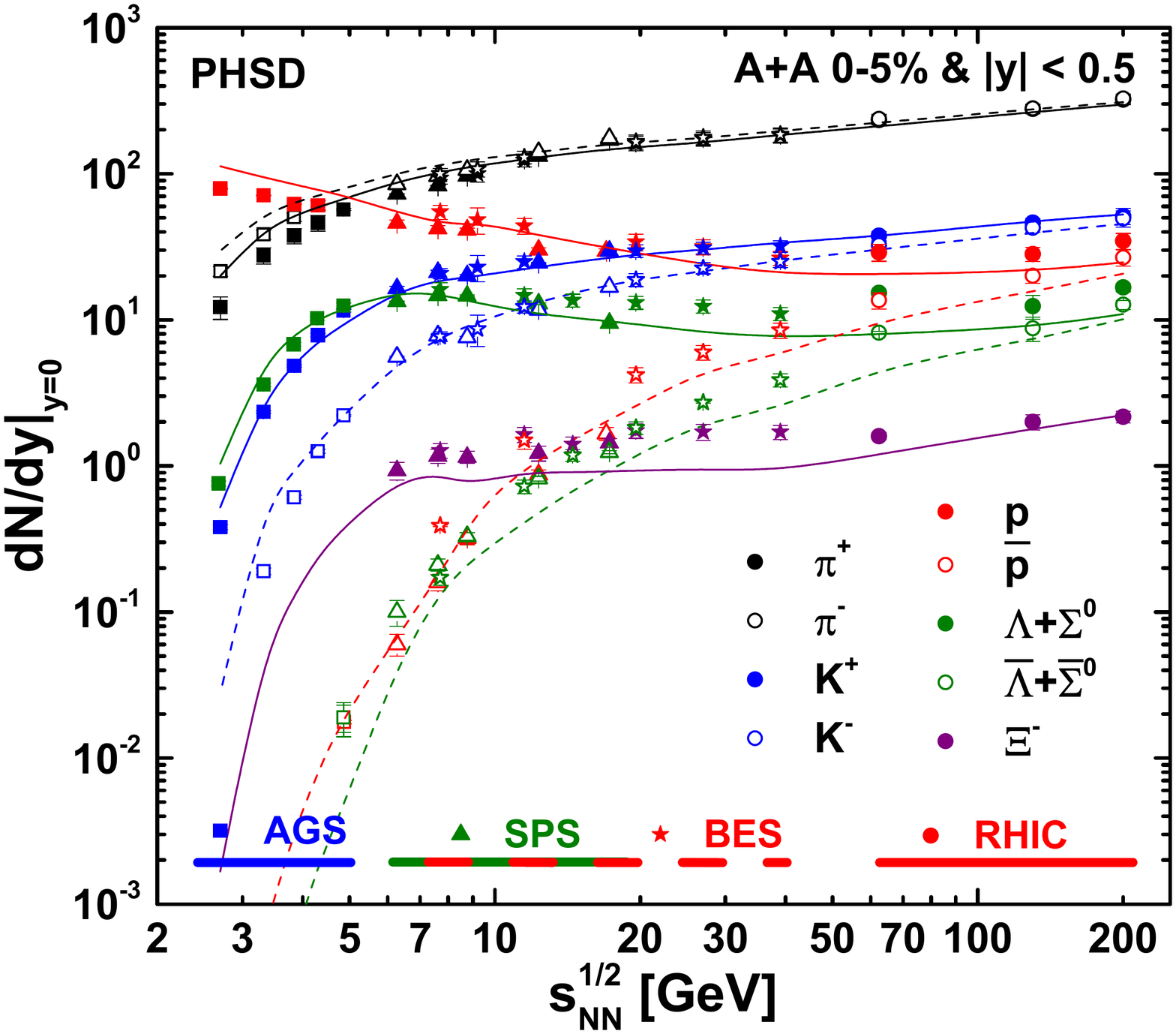}
	\caption{Excitation function $(dN/dy)|_{y=0}$ of particles produced in central heavy-ion collisions at midrapidity from the PHSD \citep{Moreau:2019aux} in comparison with experimental data from  E895 \citep{Klay:2003zf,Klay:2001tf}, E866-E917 \citep{Ahle:1999uy,Ahle:2000wq,Back:2001ai}, E896 \citep{Albergo:2002tn} and E895 \citep{Pinkenburg:2001fj} collaborations at AGS energies, from the NA49 \citep{Alt:2007aa,Afanasiev:2002mx,Alt:2006dk,Alt:2008qm,Alt:2004kq} and the NA57 \citep{Antinori:2004ee} collaborations for SPS energies, and from the STAR collaboration for the Beam-Energy-Scan \citep{Adamczyk:2017iwn,Ashraf:2016fjl} and top RHIC regimes \citep{Adcox:2002au,Abelev:2008ab,Aggarwal:2010ig}.}
	\label{fig_dNdy}
\end{figure}

In Fig. \ref{fig_dNdy} we present the excitation function of hadron multiplicities at midrapidity $dN/dy|_{y = 0}$ as a function of the collisional energy $\sqrt{s_{NN}}$. 
The symbols correspond to the experimental data, while the lines correspond to  the PHSD5.0 calculations including $(T,\mu_B)$ for mesons - $\pi^\pm, \ K^\pm$ and 
(anti-)baryons - $p, \ \bar p, \ \Lambda+\Sigma^0, \ \bar\Lambda+\bar\Sigma^0, 
\ \Xi^-$. One can see that the  mesons and  (anti-)baryons multiplicities
grow rapidly with increasing energy.
The production of particles at high energies is practically identical with respect to particles-antiparticles and mainly originates from the hadronization process of the Quark-Gluon Plasma. With decreasing energy one can see a separation between baryons and antibaryons which comes from baryon stopping, from possible $B-\bar{B}$ annihilations, and from the fact that antibaryons (especially multi-strange baryons) are preferentially produced by the hadronization process from the QGP and, thus, more sensitive to the
$(T,\mu_B)$ properties of the QGP partons.

\subsection{Directed and elliptic flows}

As follows from hydrodynamical calculations ~\citep{Bernhard:2016tnd,Marty:2013ita,Romatschke:2007mq,Song:2008si}  and 
the Bayesian analysis~\citep{Bernhard:2016tnd}, 
the results for the flow harmonics $v_n$ are sensitive to the transport coefficients
Recently we investigated the sensitivity of the collective flow ($v_1$, $v_2$) coefficients to the $\mu_B$ dependences of partonic cross sections of DQPM 
which provide the $\eta/s$ and $\zeta/s$ consistent with lQCD results as discussed in Section 2.
For that we calculated  $v_1$, $v_2$  for different identified hadrons -
mesons and (anti-) baryons \citep{Soloveva:2020xof,Soloveva:2020ozg}.

\begin{figure}[t]
\begin{center}
\includegraphics[width=7.5cm]{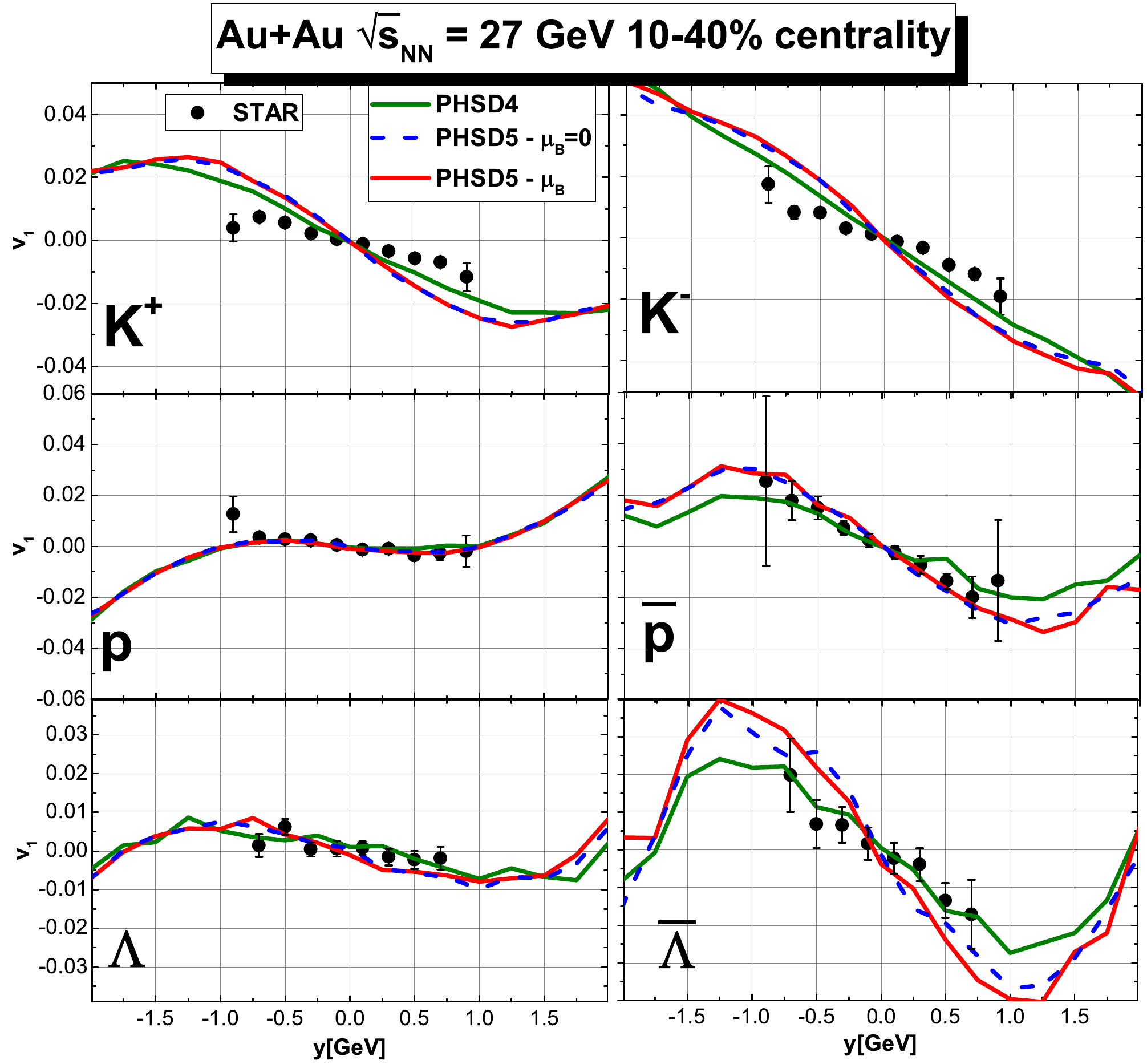}
\end{center}
\caption{Directed flow of identified hadrons as a function of rapidity for Au+Au collisions at $\sqrt{s_{NN}}$ = 27 GeV for PHSD4.0 (green lines), PHSD5.0 with partonic cross sections and parton masses calculated for $\mu_B$ = 0 (blue dashed lines) and with cross sections and parton masses evaluated at the actual chemical potential $\mu_B$ in each individual space-time cell (red lines) in comparison to the experimental data of the STAR Collaboration~\citep{Adamczyk:2017nxg}.}
\label{v1y27GeV}
\end{figure}

We start with presenting the results of our study \citep{Soloveva:2020xof,Soloveva:2020ozg}
for the traces of $\mu_B$ dependences of the QGP interaction cross sections in 
the directed flow. 
Fig. \ref{v1y27GeV} depicts the directed flow $v_1$ of identified hadrons
($K^\pm, p, \bar p, \Lambda+\Sigma^0, \bar\Lambda + \bar\Sigma^0$) versus
rapidity for Au+Au collisions at $\sqrt{s_{NN}}$ = 27 GeV.
One can see a good agreement between PHSD results and experimental data from the STAR collaboration~\citep{Adamczyk:2017nxg}.
However, the different versions of PHSD for the $v_1$ coefficients show a quite
similar behavior; only antihyperons indicate a slightly different flow.
This supports again the finding that strangeness, and in particular anti-strange
hyperons, are the most sensitive probes for the QGP~properties.

\begin{figure}[h]
\centering
\includegraphics[width=8cm]{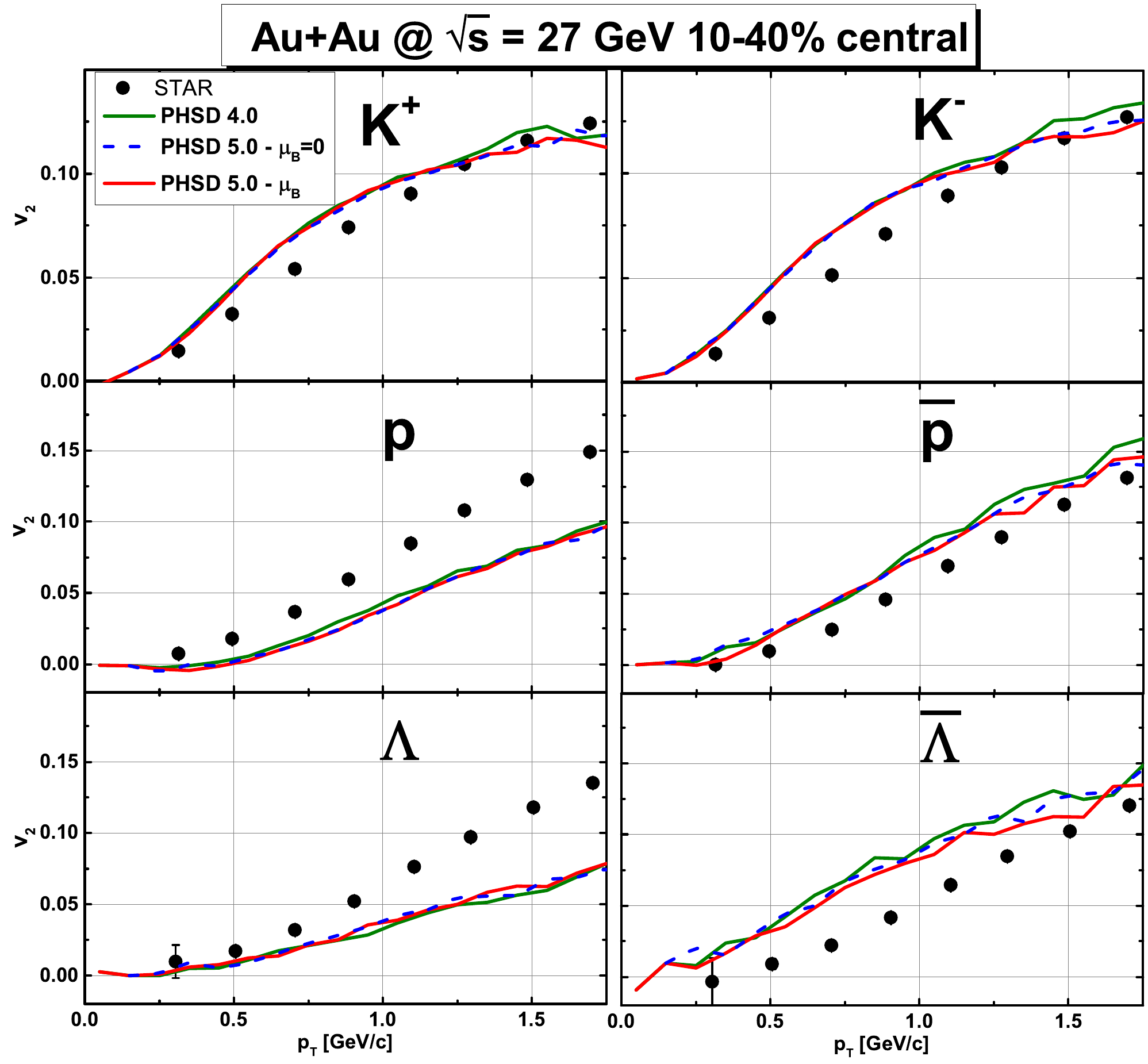}
\caption{Elliptic flow of identified hadrons ($K^\pm, p, \bar p, \Lambda+\Sigma^0, \bar\Lambda + \bar\Sigma^0$) as a function of $p_T$ for Au+Au collisions at $\sqrt{s_{NN}}$ = 27~GeV  for PHSD4.0 (green lines), PHSD5.0 with partonic cross sections and parton masses calculated for $\mu_B$ = 0 (blue dashed lines) and with cross sections and parton masses evaluated at the actual chemical potential $\mu_B$ in each individual space-time cell (red lines)
in comparison to the experimental data of the STAR Collaboration~\citep{Adamczyk:2015fum}.}
\label{v2y27GeV}
\end{figure}

We continue with presenting the results for the elliptic flow of charged hadrons
from heavy-ion collisions within the PHSD5.0.
In Fig. \ref{v2y27GeV} we display the elliptic flow $v_2$ of identified hadrons 
($K^\pm, p, \bar p, \Lambda+\Sigma^0, \bar\Lambda + \bar\Sigma^0$) 
as a function of  $p_T$ at $\sqrt{s_{NN}}$ = 27 GeV  for PHSD4.0 (green lines), 
PHSD5.0 with partonic cross sections and parton masses calculated for $\mu_B = 0$
(blue dashed lines) and with cross sections and parton masses evaluated at
the actual chemical potential $\mu_B$ in each individual space-time cell (red lines) 
in comparison to the experimental data of the STAR Collaboration~\citep{Adamczyk:2015fum}.
Similar to the directed flow shown in Fig. \ref{v1y27GeV}, the elliptic flow from all
three scenarios in the PHSD shows a rather similar behavior;  
the differences are very small (within the statistics achieved here). 
Only antiprotons and antihyperons show a tendency for a small decrease of the
$v_2$ at larger $p_T$ for PHSD5.0 compared to PHSD4.0, which can be attributed
to the explicit $\sqrt{s}$-dependence and different angular distribution of partonic cross sections in the PHSD5.0. 
We note that the underestimation of $v_2$ for protons
and $\Lambda$'s might be attributed to details of the hadronic vector potentials involved  in this calculations which seem to underestimate the baryon repulsion.
We also attribute the slight overestimation of antibaryon $v_2$ to the lack of baryon potential.

\begin{figure}[h]
\centering
\includegraphics[width=8.5cm]{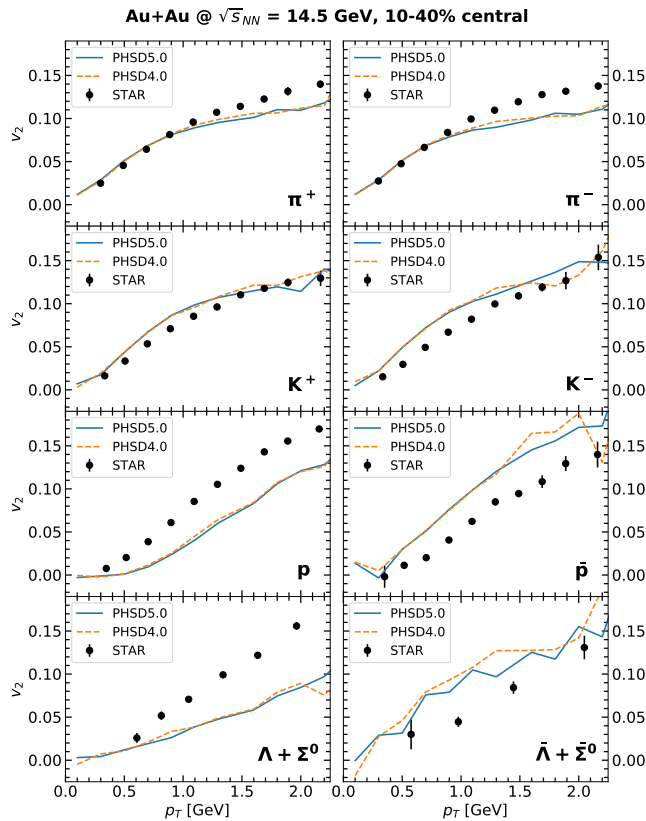}
\caption{Elliptic flow of identified hadrons ($\pi^\pm, K^\pm, p, \bar p, \Lambda+\Sigma^0, \bar\Lambda + \bar\Sigma^0$) as a function of $p_T$ for Au+Au collisions at $\sqrt{s_{NN}}$ = 14.5~GeV  for  the PHSD4.0 (orange dashed lines) and the PHSD5.0 with  cross sections and parton masses 
depending on $\mu_B$ (blue solid lines) in comparison to the experimental data of the 
STAR Collaboration~\citep{Adam:2019dkq}.}
\label{v2pt14GeV}
\end{figure}

In this study we also explore the lower collision energy $\sqrt{s_{NN}}$ = 14.5 GeV 
and present in Fig. \ref{v2pt14GeV} the results for the $v_2$ of identified hadrons
($\pi^\pm, K^\pm, p, \bar p, \Lambda+\Sigma^0, \bar\Lambda + \bar\Sigma^0$) as a function of $p_T$. We find the same tendency as in Fig. \ref{v2y27GeV}
and show explicitly the PHSD4.0 (orange dashed lines) and the PHSD5.0 calculations  
with cross sections and 
parton masses depending on $\mu_B$ (blue solid lines) 
in comparison to the experimental data of the STAR Collaboration~
\citep{Adam:2019dkq}.

\section{Conclusions}

In this study we have reviewed the thermodynamic properties of the QGP as created in heavy-ion collisions and reported on the influence of the baryon chemical potential 
$\mu_B$ on the experimental observables such as the directed and elliptic flow.

For the description of the QGP we have employed the extended 
Dynamical QuasiParticle Model (DQPM) that is matched to reproduce the lQCD equation-of-state versus temperature $T$ at zero and finite baryon chemical potential $\mu_B$.
We recall that the ratios $\eta/s$ and $\zeta/s$ from the DQPM agree very well with
the lQCD results from Ref.~\citep{Astrakhantsev:2018oue} and show a similar behavior
as the ratio obtained from a Bayesian fit~\citep{Bernhard:2016tnd}. As found previously in Refs. \citep{Moreau:2019vhw,Soloveva:2019xph} the transport coefficients show 
only a mild dependence on $\mu_B$.

Our study of the heavy-ion collisions, i.e. non-equilibrium QGP,
has been performed within the extended Parton--Hadron--String Dynamics (PHSD)
transport approach ~\citep{Moreau:2019vhw} in which the properties of quarks 
and gluons, i.e. their masses and widths, depend on $T$ and $\mu_B$ explicitly.
Moreover, the partonic interaction cross sections are obtained
by evaluation of the leading order Feynman diagrams from the DQPM effective 
propagators and couplings  and explicitly depend on
the invariant energy $\sqrt{s}$ as well as $T$ and $\mu_B$.

We have demonstrated the time evolution of heavy-ion collisions and
$T-\mu_B$ trajectories as obtained from the PHSD5.0 calculations for SPS energies.
We have explored the energy dependence of the averaged $\mu_B$ around 
the chemical freeze-out temperature and found a reasonable agreement with statistical
model results \citep{Andronic:2009jd,Cleymans:2005xv,Adamczyk:2017iwn}. 

We have explored the sensitivity of heavy-ion observables such as the
excitation function of the hadron multiplicities and
collective flow coefficients $v_1, v_2$ on the ($T,\mu_B$) dependence of the QGP properties.
After remanding  that the sensitivity (w.r.t. the $\mu_B$-dependence) of hadronic rapidity and $p_T$ distributions of hadrons for symmetric and asymmetric heavy-ion collisions from AGS to RHIC energies was found to be very low \citep{Moreau:2019vhw,Soloveva:2019xph}, we have discussed  the sensitivity of the collective  flow of hadrons.  As an example, we have shown the results for 
$\sqrt{s_{NN}}$ = 14.5 and 27~GeV .
As has been shown in Refs. \citep{Soloveva:2020xof,Soloveva:2020ozg} we find only very small differences between the results from PHSD4.0 and from PHSD5.0 
on the hadronic flow observables at high as well as at intermediate energies.  This is related to the fact that at high energies, where the matter is dominated by the QGP,
one probes only a small baryon chemical potential in central collisions at midrapidity,
while at lower energies (and larger $\mu_B$) the fraction of the QGP drops rapidly
(cf. Fig. \ref{Fig_QGPfrac}) such that in total the final observables are
dominated by the hadronic interactions and thus the information about the partonic properties and scatterings is washed out. 
This finding is also consistent with the fact that the transport coefficients -
as evaluated in the DQPM - show only a weak dependence on $\mu_B$.
We have shown that the mild $\mu_B$-dependence of QGP interactions is more
pronounced in observables for strange hadrons (kaons and especially
anti-strange hyperons) which provides an experimental hint for the search of  
$\mu_B$ traces of the QGP for experiments at the future FAIR/NICA facilities 
and the BESII program at RHIC.

\section*{Acknowledgements}
The authors acknowledge inspiring discussions with J\"org Aichelin, 
Wolfgang Cassing, Ilya  Selyuzhenkov and Arkadiy Taranenko. 
The computational resources have been provided by
the LOEWE-Centerfor Scientific Computing and the "Green Cube" at GSI, Darmstadt.
P.M. acknowledges support by the U.S. D.O.E. under Grant No. 
DE-FG02-05ER41367.
O.S., I.G.  acknowledge  support  from HGS-HIRe  for  FAIR. 
L.O. has been financially supported  by the Alexander von Humboldt Foundation.
Furthermore, we acknowledge support by the Deutsche Forschungsgemeinschaft 
(DFG, German Research Foundation): grant BR 4000/7-1, grant
CRC-TR 211 ’Strong-interaction matter under extreme conditions'  -  Project number  315477589  -  TRR  211; by the Russian Science Foundation grant 19-42-04101;
by the European Union’s Horizon 2020 research and innovation program under grant agreement No 824093 (STRONG-2020) and by the COST Action THOR, CA15213.

\bibliography{Biblio_brat}%

\end{document}